\documentclass[aps,prf,reprint,groupedaddress]{revtex4-2}

\usepackage{graphicx}% Include figure files
\usepackage{dcolumn}% Align table columns on decimal point
\usepackage{bm}% bold math
\usepackage[utf8]{inputenc}
\usepackage[T1]{fontenc}
\usepackage{mathptmx}
\begin{document}

\title{Energy cascades in active-grid-generated turbulent flows}


\author{D.O. Mora}
%\thanks{}
\altaffiliation{Department of Mechanical Engineering, University of Washington, Seattle, Washington 98195-2600, USA}
\author{E. Mu\~niz Pladellorens}
\author{P. Riera Turr\'o}
\author{M. Lagauzere}
\author{M. Obligado}
 \email{martin.obligado@univ-grenoble-alpes.fr}

\affiliation{ 
Univ. Grenoble Alpes, CNRS, Grenoble INP, LEGI, 38000 Grenoble, France%\\This line break forced with \textbackslash\textbackslash
}%


\date{\today}

\begin{abstract}

The energy cascade and diverse turbulence properties of active-grid-generated turbulence were studied in a wind tunnel via hot-wire anemometry. To this aim, two active grid protocols were considered. The first protocol is the standard triple-random mode, where the grid motors are driven with random rotation rates and directions, which are changed randomly in time. This protocol has been extensively used due to its capacity to produce higher values of $Re_\lambda$ than its passive counter part, with good statistical homogeneity and isotropy. The second protocol was a static or open grid mode, where all grid blades were completely open, yielding the minimum blockage attainable with our grid.

Centreline streamwise profiles were measured for both protocols and several inlet velocities. It was found that the turbulent flow generated with the triple-random protocol evolved in the streamwise direction consistently with an energy dissipation scaling of the form $\varepsilon=C_\varepsilon u^{\prime3}/L$, with $C_\varepsilon$ being a constant, $L$ the longitudinal integral length-scale, and $u^\prime$ the rms of the longitudinal velocity fluctuations.

Conversely for the open-static grid mode, the energy dissipation followed a non-equilibrium turbulence scaling, namely, $C_\varepsilon \sim Re_G/Re_L$, where $Re_G$ is a global Reynolds number based on the inlet conditions of the flow, and $Re_L$ is based on the local properties of the flow downstream the grid. Furthermore, this open-static grid mode scaling exhibits important differences with other grids, as the downstream location of the peak of turbulence intensity is a function of the inlet velocity, a remarkable observation not previously reported, as it would allow to study the underlying principles of the transition between equilibrium and non-equilibrium scalings, which are yet to be understood. 

It was also found that a rather simple theoretical model can predict the value of $C_\varepsilon$ based on the number density of zero-crossings of the longitudinal velocity fluctuations. A theory, which is valid for both active grid operating protocols (and therefore two different energy cascades).

\end{abstract}

%\keywords{}

\maketitle

\section{Introduction}

Since the first active grid was proposed by Makita~\cite{makita1991active}, active-grid-generated turbulence has become a standard way to generate moderate-to-high Reynolds numbers in wind/water tunnels~\cite{mydlarski2017turbulent}. They present several advantages for wind/water tunnels research, as they allow to obtain bespoke unstationary/inhomogeneous turbulence and generate large values of Reynolds numbers based on the Taylor micro-scale $Re_\lambda$ while keeping reasonable homogeneous isotropic turbulence (HIT) conditions. Their use has therefore become widespread in several active research fields, such as two-phase~\cite{obligado2014preferential, prakash2012gravity} and  atmospheric flows~\cite{knebel2011atmospheric}, wind energy~\cite{cal2010experimental} and fundamental turbulence~\cite{mordant2008experimental, sinhuber2015decay, mydlarski1996onset}.

In many of these applications, these grids are used to `tune' the turbulence intensity and/or the Reynolds number $Re_\lambda$. Several grid protocols are then available to explore the parameter space, but little or no attention has been paid yet to the consequences of them on the energy cascade. For instance, some studies have focused on the characteristics of the turbulent flow for a given blade geometry and initial conditions~\cite{hearst2015effect}. Furthermore, other studies have reported values of the turbulent kinetic energy dissipation constant $C_\varepsilon$~\cite{thormann2014decay, puga2017normalized}, basically employing one operating protocol with subtle variations (the one defined as triple random below), and with the intent to corroborate the validity of the scalings derived from Kolmogorov 1941 theory, also detailed below. $C_\varepsilon$ is a key parameter to understand these flows, as it can be used to determine the properties of the energy cascade on them (for further details, we refer to the review from Vassilicos~\cite{vassilicos2015dissipation}).

Therefore, it remains unclear whether the nature of the energy cascade remains unchanged if  strong changes were to be introduced in the grid controlling protocol. The traditional view, derived from Kolmogorov scalings (and compatible with both his 1941 and 1962 theories), is that the turbulent kinetic energy dissipation follows the law $\varepsilon=C_\varepsilon \frac{u^{\prime3}}{L}$, where $u^\prime$ is the standard deviation of streamwise velocity fluctuations $u$ and $L$ the longitudinal integral length-scale. $C_\varepsilon$ is a constant that may depend on the boundary conditions but remains constant for a fixed grid geometry. Recently, different studies have reported that these scalings may also be fulfilled within a balanced-non equilibrium cascade~\cite{goto2016unsteady}. From now on, we will refer to these scalings as `standard' dissipation scalings.

It has been known for some time that grids can generate a region where turbulence is at odds with these laws, i.e., $C_\varepsilon$ is not constant, but instead it goes as $C_\varepsilon \sim Re_G^m/Re_L^n$, where $Re_G$ is a Reynolds number that depends on the inlet conditions, and $Re_L$ is a local, streamwise position dependent one~\cite{vassilicos2015dissipation}. The exponents $n$ and $m$ have been found to be very close to unity; $m=n=1$ for large values of $Re_\lambda$. For grid turbulence, $Re_G=MU_\infty/\nu$, with $M$ being the mesh spacing, $U_\infty$ the inlet velocity and $\nu$ the kinematic viscosity of the flow. The local Reynolds number is defined as $Re_L=Lu^\prime/\nu$. These high Reynolds non-equilibrium scalings have been recovered in
both regular and fractal grids, at a range of downstream positions
which starts close to the peak of turbulent kinetic energy and extends
well beyond it. Typically, this region spans $x_{max}<x<5x_{max}$, where $x$ is the downstream coordinate from the grid, and $x_{max}$ (see details below) can be predicted as the position where the wakes of the grid bars meet, which hinge on the grid geometry.
 
The presence of different energy cascades would create limitations for the applicability of active grids in some situations, as their respective consequences regarding scales separation, and the number of degrees of freedom (quantified by $L/\lambda$), would follow different laws. `Standard' scalings verify the relation  $L/\lambda \sim C_\varepsilon Re_\lambda$, while high Reynolds non-equilibrium scalings preserve $L/\lambda \sim \sqrt{Re_G}$ (where both formulae assume $\varepsilon \sim \nu u'^2/\lambda^2$). The latter being independent of $Re_\lambda$ at fixed $Re_G$.

This work summarizes a series of experiments on active-grid-generated turbulence. We found evidence that different operating protocols produce different energy cascades. Given the myriad of possible operating protocols, we focused on two extreme cases, one where the grid was static and fully opened, and the other where the grid moves randomly (via the triple random protocol, explained in the next section). Finally, we show that it is possible to check the consistency of the values of $C_\varepsilon$ obtained via an adaptation of the Rice theorem to turbulent flows.

\section{Experimental set-up}

\begin{figure}[t!]
\centering
\includegraphics[width=0.4\textwidth]{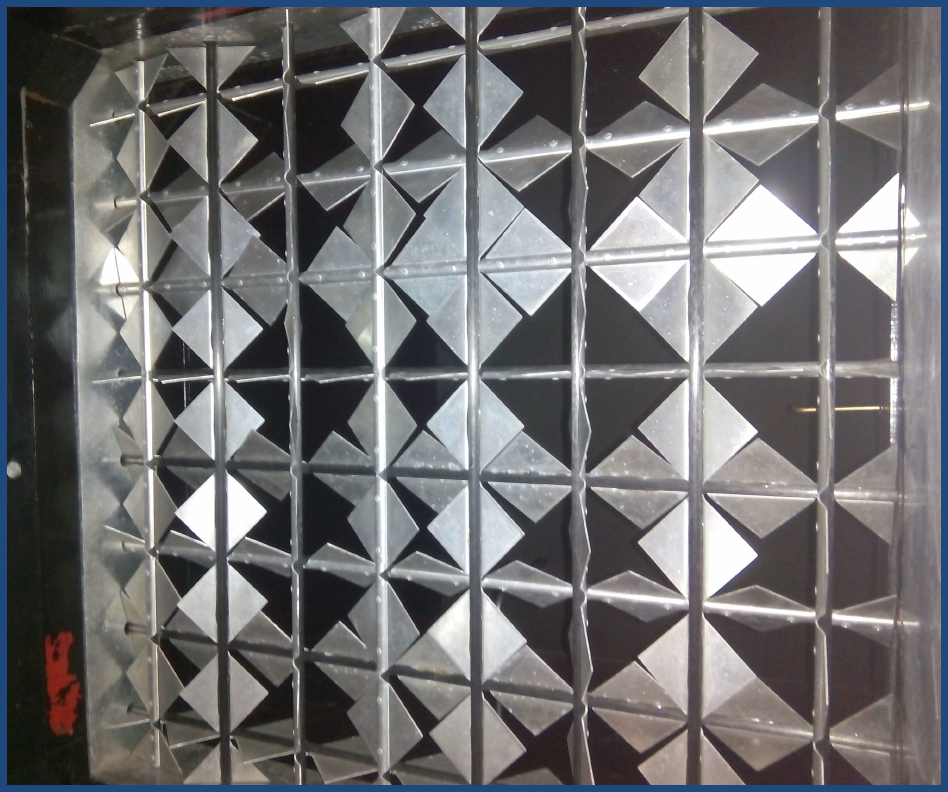}
\caption{Picture of the active grid used on the present work.}
\label{fig1}
\end{figure}

Experiments were conducted in the Lespinard wind tunnel at LEGI: a large wind tunnel with a measurement section of $4$~m long and a square cross-section of $0.75\times0.75$~m$^2$. Turbulence is generated with an active grid made of 16 rotating axes (eight horizontal and eight vertical, with a mesh size $M=10$~cm) mounted with co-planar square blades (with also a 10~cm side, see figure \ref{fig1}). Each axis is driven independently with a step motor whose rotation rate and direction can be changed dynamically. Two different protocols were tested; one where the blades are moving randomly and one where they remain static. For the random mode (hereafter referred as active grid or AG), the motors were driven with a random rotation rates and directions, which were changed randomly in time (the velocity was varied between 1 and 3 Hz, and changes for a lapse were between 1 and 3s). This mode is of widespread use to generate moderate-to-high $Re_\lambda$ with good HIT conditions and is usually called the triple random mode. The second protocol employed was the static open mode (referred as open grid or OG), where the grid was completely open (thus minimising blockage), and static (and therefore generating low values of $Re_\lambda$). More details on the active-grid and the wind tunnel can be found in a previous work~\cite{obligado2014preferential}.

All measurements were made by means of a single hot wire, using a Dantec Dynamics 55P01 hot-wire probe, driven by a
Dantec StreamLine CTA system. The Pt-W wires were $5 \mu m$ in diameter, $3$ mm long with a sensing length of $1.25$ mm. Acquisitions were made for $300$ s at $25$ kHz and $50$ kHz (while a low-pas filter was always set-up at 30 kHz to counteract for aliasing). It was checked that for all the datasets where $C_\varepsilon$ results are reported we have at least $\kappa \eta = \frac{2 \pi }{ U} f \eta =1$ (with $U$ the local mean velocity).

For each operating protocol, two different kinds of measurements were done: streamwise profiles at constant inlet velocity $U_\infty$, and measurements at fixed streamwise positions ($x=150$ and $x=300$cm) and varying $U_\infty$. The range of $U_\infty$ explored is imposed by the inherent instabilities of the wind tunnel at low velocities, and by the fact that at high $U_\infty$ the grid motors do not have enough power to compensate for the drag force of incoming wind, i.e., impairing the randomness of the active grid protocol and closing the open grid, respectively.

For AG (the active grid protocol), one streamwise profile at fixed inlet velocity was recorded at $U_\infty=6.7$m/s between $x=45$ cm, and $x=325$ cm. Next, profiles at fixed downstream position were taken at 5 different values of $U_\infty$ (1.8, 2.6, 3.6, 5.2 and 6.8 m/s). 

For OG (the open grid protocol), three  streamwise profiles at fixed inlet velocity were recorded on a similar range as AG ($U_\infty=$8.6, 11.9 and 17.0 m/s). Likewise, the profiles for the fixed downstream position were taken at 5 different values of $U_\infty$ (4.4, 7.0, 9.7, 12.2 and 14.7 m/s). 

\begin{figure}[t!]
\centering
\includegraphics[width=0.5\textwidth]{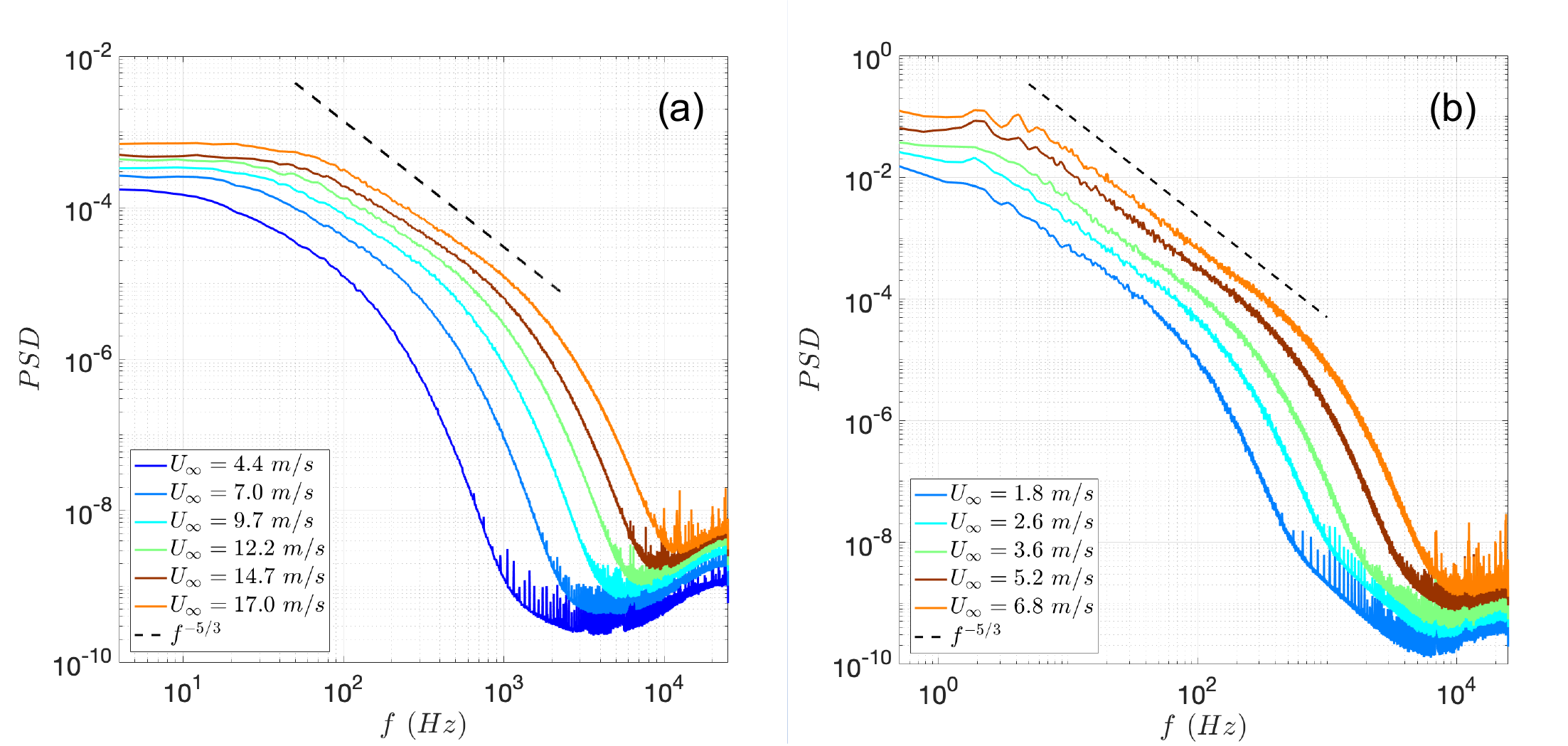}
\caption{Power spectral densities of velocity fluctuations obtained at $x=300$ cm for different values of $U_\infty$ for the open (a) and the active (b) modes. The black dashed line is a $-5/3$ power law.}
\label{fig1a}
\end{figure}

The registered time signals were subsequently converted into the spatial domain via the Taylor hypothesis. To account for bias of this method, a modified Taylor hypothesis that takes into account a local mean velocity of the flow~\cite{pinton1994correction} was also checked, obtaining almost identical results in both cases. Figure \ref{fig1a} shows the velocity fluctuations power spectral density obtained for the whole range of velocities -and modes- at the downstream position $x=300$cm. It can observed that they all exhibit a power-law close to a $-5/3$. Nevertheless, at low values of $U_\infty$, the OG has a less clear $5/3$ exponent, but its shape still remains very similar to other regular static grid spectra previously reported ~\cite{antonia2014collapse}.

\begin{figure}[t!]
\centering
\includegraphics[width=0.5\textwidth]{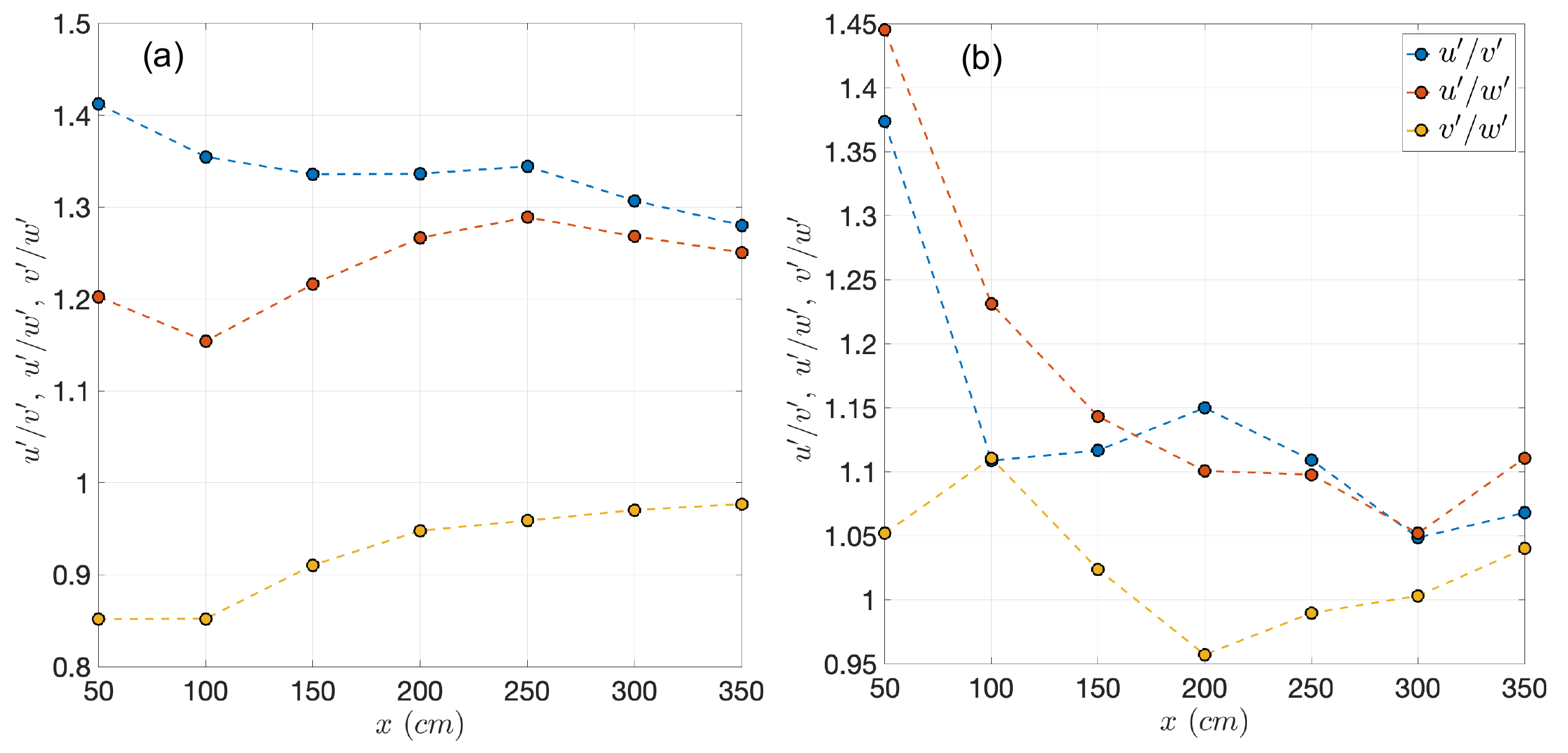}
\caption{Ratios between the fluctuating velocity vector $\vec{U^\prime}={\left(u,v,w\right)}$ components for the open (left) and the active (right) modes. Measurements were taken at $U_\infty=11.5$~m/s for the first and at $U_\infty=4.9$~m/s for the latter modes.}
\label{figiso}
\end{figure}

Large scale isotropy was quantified with a COBRA probe manufactured by TFI, which is able to compute the three fluctuating velocity components with a resolution of $\sim 500$Hz. Figure \ref{figiso} shows the ratios between the three components of the fluctuating velocity vector $\vec{U^\prime}={\left( u,v,w \right)}$. The active mode  exhibits acceptable isotropy conditions, with ratios below 10\% for moderate distances away from the grid, i.e., $x>1$m. Surprisingly, the isotropy ratios for the open mode are larger (in the order of 30\%), and consistent with values reported for fractal grids in the range which comprises non-equilibrium turbulence~\cite{hurst2007scalings}. Hence, it is seen that reasonable isotropy conditions are found on both modes. However, caution has to be taken when analyzing results coming from the open mode.

\begin{figure*}[t!]
\centering
\includegraphics[width=0.95\textwidth]{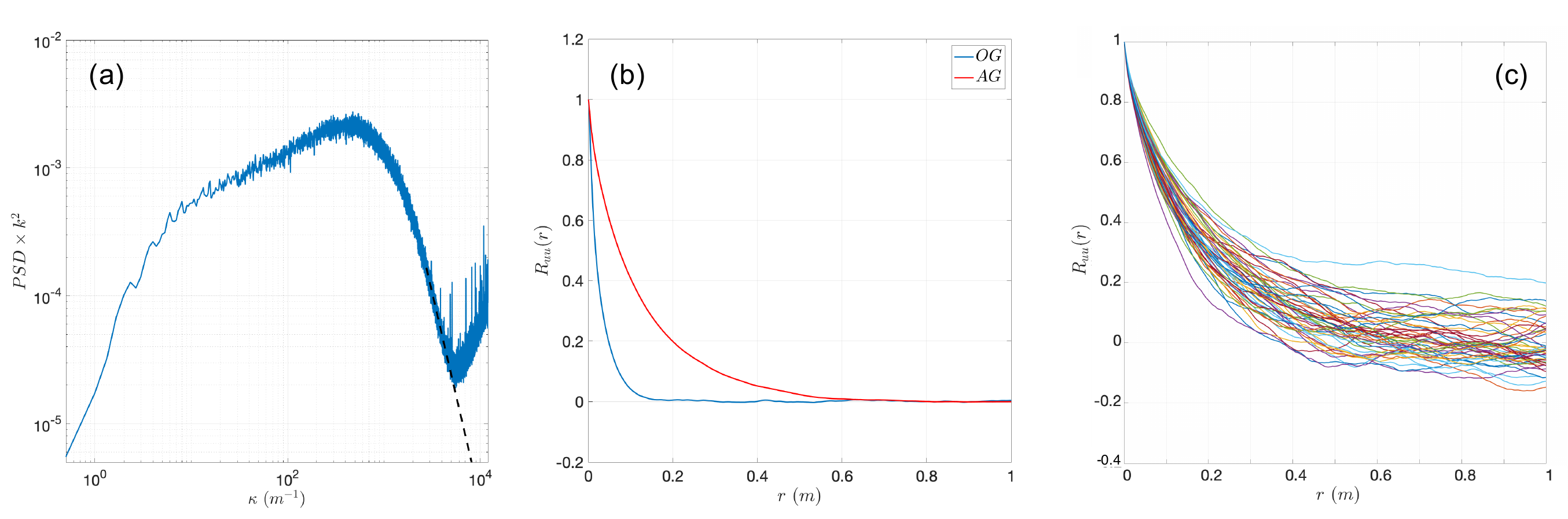}
\caption{ (a) Estimation of $\varepsilon$ via the dissipation spectra; the black dashed line corresponds to the modelled frequencies. (b) Typical autocorrelation functions $R_{uu}$ obtained for both operating modes. (c) Example of the method applied to estimate $L$ for the active mode.}
\label{fig2}
\end{figure*}

The turbulent dissipation rate $\varepsilon$ was estimated via the dissipation spectrum. It was calculated as $\varepsilon=\int 15 \nu
k_{1}^2 E_{11}dk_{1}$ where $E_{11}(k_{1})$ is the 1D power
spectrum, $k_{1} =2\pi f/U$ is the respective wave number, and $f$ is the Fourier frequency in
Hz. The latter involves assuming local, small-scale, isotropy, and applying the Taylor hypothesis. The noise at high frequencies has been removed and modeled as a power law, fitted for each time signal (figure \ref{fig2}a). The Taylor micro-scale has been obtained from $\varepsilon$ as $\lambda=\sqrt{15 \nu <u^{2}>/\varepsilon}$.

The integral lengthscale $L$ was estimated via the streamwise velocity autocorrelation function $R_{uu}$ (figure \ref{fig2}b). For OG, it was estimated as $L=\int_0^{r_0}R_{uu}dr$, with $r_0$ the smallest value at which $R_{uu}=0$, and $r$ estimated again via the Taylor hypothesis.

The estimation of $L$ for the AG is more difficult, as the autocorrelation function has the pitfall of not always crossing zero. Thus, it was estimated using a method proposed by Puga \& LaRue~\cite{puga2017normalized}, which consists on dividing the velocity signal in small, not converged, segments, that present a typical dispersion $\delta$ in $R_{uu}$ (figure \ref{fig2}c). Therefore, $L$ can be computed as $L=\int_0^{r_\delta}R_{uu}dr$, where now $r_\delta$ is the smallest value for which $R_{uu}=\delta$. This method, however, presents some ambiguity regarding the absolute value of $L$ as it strongly depends on the length of the segments chosen to estimate $\delta$. It can be appreciated in figure \ref{fig2} that our choice of small segments yielded large values of $\delta$. This choice reduces the dispersion of $L$ between datasets, but it also reduces the value of $L$. We have examined for biases with respect to this decision, and it was found that different averages present the same trends. 

The previous discussions show that both $C_\varepsilon$ and $L$ were estimated without assuming any K41 scalings. Thus, it is possible to compute $C_\varepsilon=\varepsilon L / u^{\prime3}$ without any assumptions except for the presence of local homogeneity which would allow a cascade to operate without interference of gradients of one-point flow statistics. The range of turbulence parameters obtained for each operating protocol is shown in table \ref{tab:turbParams}.

\begin{table}[b]
\begin{center}
\begin{tabular}{|c|c|c|}
\hline
Parameter & OG & AG \\
\hline
$U_\infty~(m/s)$ & 4.4--17.0 & 1.8--6.8 \\
$u^\prime/\langle u\rangle ~(\%)$ & 2.0--10.0 & 12.5--45.0 \\
$Re_\lambda$ & 50--200 & 200--950 \\
$\lambda~(mm)$ & 3.0--8.0 & 7.0--14.0 \\
$\eta~(\mu m)$ & 100--400 & 100-500 \\
$L~(cm)$ & 1.0--3.0 & 5.0--13.0 \\
\hline
\end{tabular}
\end{center}
\caption{Typical turbulence parameters range for the open (OG) and active (AG) modes: inlet velocity $U_\infty$, turbulence fluctuations $u^\prime/\langle u\rangle$, Reynolds number based on the Taylor micro-scale $Re_\lambda$, Taylor micro-scale $\lambda$, Kolmogorov lengthscale $\eta=(\nu^3/\varepsilon)^{1/4}$ and streamwise integral length-scale $L$.}\label{tab:turbParams}
\end{table}

\section{Results}

\subsection{Active grid mode}

\begin{figure*}[t!]
\centering
\includegraphics[width=0.95\textwidth]{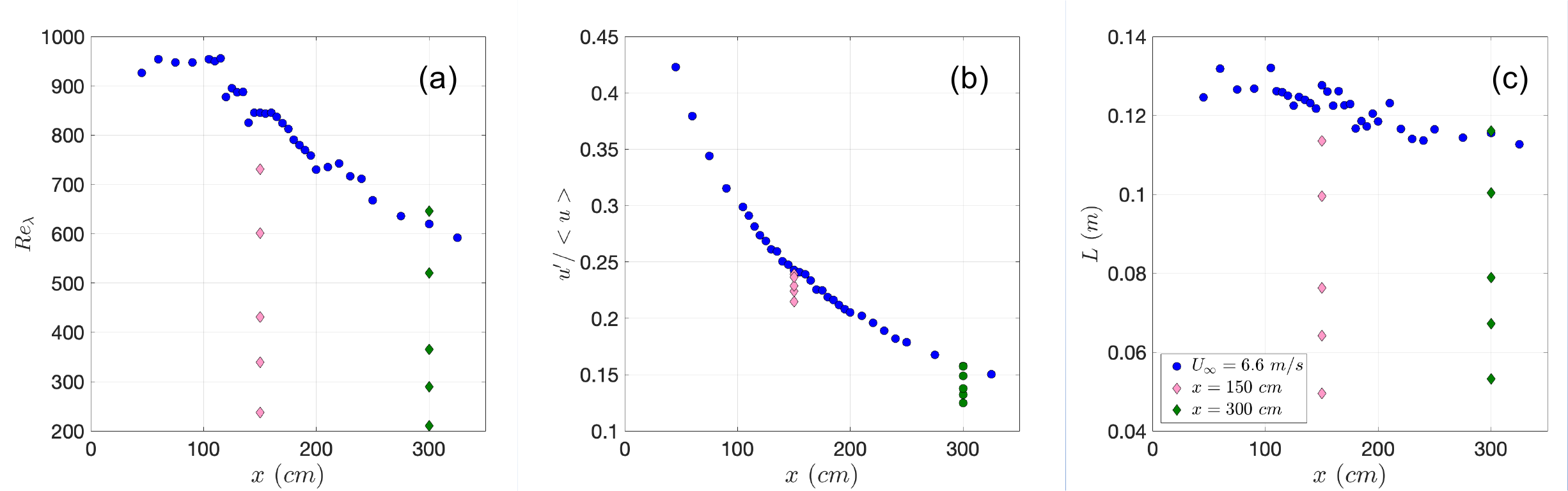}
\caption{Streamwise evolution of $Re_\lambda$ (a), turbulent fluctuations (b) and integral length-scale $L$ (c) for all results obtained for the active mode (AG).}
\label{fig3}
\end{figure*}

In this section we present results for the grid operated in the active, triple random, mode. Figure \ref{fig3} shows different turbulence parameters obtained for the measured streamwise profiles. No peak was observed for the velocity fluctuations, including previous measurements much closer to the grid ($x\sim10$~cm, not shown here). The absence of this peak can be due to the inability of the hot wire to properly resolve the velocity at the high fluctuations present very close to the grid. It can be observed that fluctuations have a considerable magnitude, and therefore the Taylor hypothesis should be used with care. We will therefore only show results in the following for $x>130$~cm, where fluctuations remain below 25\%, similar to results reported in other works~\cite{sinhuber2015decay}. However, the validity of this approach  (using the Taylor hypothesis under these conditions) remains an open question and the validity of the use of this hypothesis on active-grid-generated turbulence should be addressed in detail in future works. 

From the figures, it can be seen that first; the magnitude of $Re_\lambda$ is considerable, and it significantly changes during the downstream evolution of the flow, being the latter an important requirement to disentangle standard from non-equilibrium energy cascades scalings. Secondly, the length-scale $L$ estimated via the method detailed on the previous section is smooth and slightly decreases with $x$ (a surprising result but similar trends for $L$ have been reported by Thormann \& Meneveau for fractal active grids at large $x/M$~\cite{thormann2014decay}). Finally, all parameters from figure \ref{fig3} are quite sensitive to $U_\infty$, even the turbulent fluctuations (that usually remain constant at fixed $x$ for static grids) and $\lambda$ and $\eta$ (not shown in the figure). This phenomenon has also already been reported in a previous work~\cite{hearst2015effect}. 

\begin{figure}[t!]
\centering
\includegraphics[width=0.5\textwidth]{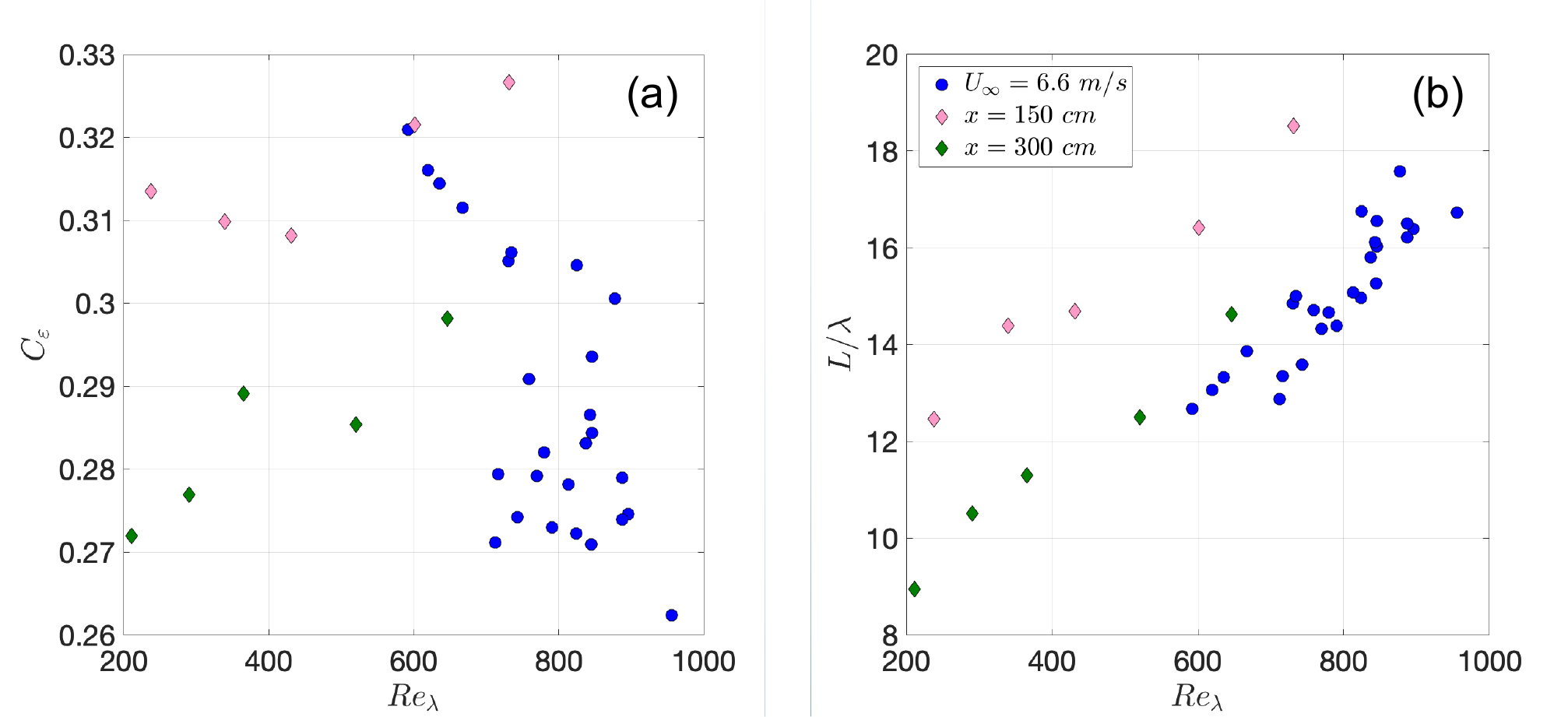}
\caption{(a) Plot of $C_\varepsilon$ vs $Re_\lambda$. (b) $L/\lambda$ vs $Re_\lambda$. Results correspond to all datasets obtained for the active mode.}
\label{fig4}
\end{figure}

The properties of the energy cascade were studied for $x>1.3m$. The figure \ref{fig4}a shows $C_\varepsilon$ vs $Re_\lambda$. There, it can be observed that for all conditions studied $C_\varepsilon$ remains constant, with a mean value of $\sim 0.3$. The latter is consistent with standard dissipation scalings, and with the relation $L/\lambda \sim C_\varepsilon Re_\lambda$, that also matches our data (figure \ref{fig4}b). The robustness of this outcome validates the use of the Taylor hypothesis in this study, and the assumption that the turbulent flow is approximately close to HIT conditions
 
Our results, however, differ from those from Puga \& LaRue~\cite{puga2017normalized} in two ways. First, we do not observe any variation of $C_\varepsilon$ with $Re_\lambda$ (while in the referred work the correlation $C_\varepsilon=2e^{-0.0108Re_\lambda}+0.647$ is proposed). Secondly, we find smaller values of $C_\varepsilon$. The discrepancy in the magnitudes of  $C_\varepsilon$ can be attributed to the underestimation of $L$, as previously detailed. Thus, an alternative method that allows to check the validity, and consistency of the values is required. This will be covered in the section \ref{sec:zeros}. 

\subsection{Open grid mode}

\begin{figure*}[t!]
\centering
\includegraphics[width=0.95\textwidth]{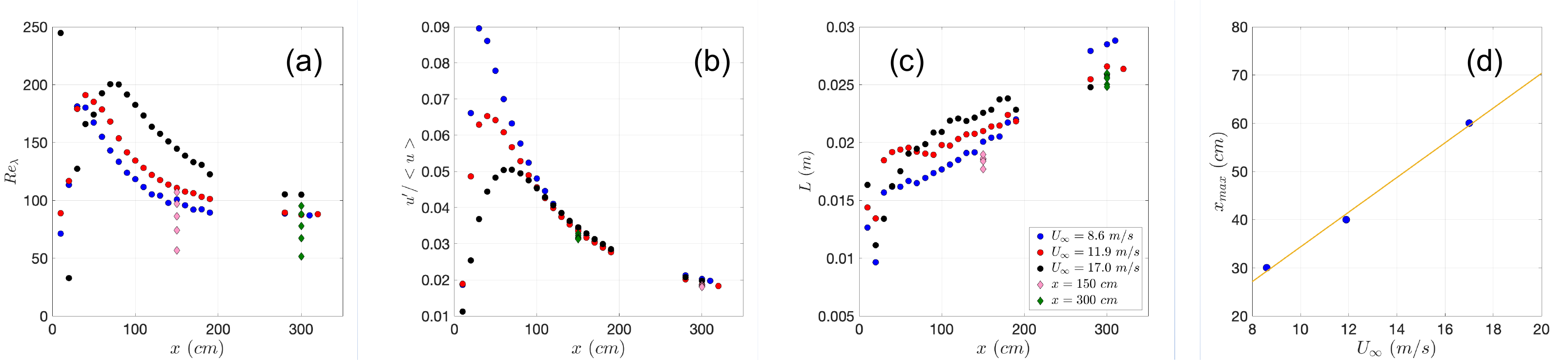}
\caption{Streamwise evolution of $Re_\lambda$ (a), turbulent fluctuations (b) and integral length-scale $L$ (c) for all results obtained for the open mode. (d) downstream position of the turbulence intensity maximum $x_{max}$ vs $U_\infty$.}
\label{fig5}
\end{figure*}

The results for the grid operated in open static mode (OG) are presented below. This case presents important differences when compared with the AG, for instance, lower values of $Re_\lambda$ (figure \ref{fig5}a) and $L$, and much lower values of turbulent fluctuations. Interestingly, and contrary to the AG, a peak in the downstream evolution of the intensity of the turbulent fluctuations was captured.

The location of the peak could be obtained via the wake interaction length $x_\star$, proposed by Mazellier \& Vassilicos~\cite{mazellier2010turbulence} and refined by Gomes-Fernandes and colaborators~\cite{gomes2012particle} (for both regular and fractal grids).  The model assumes that the maximum of turbulence intensity is a consequence of the interaction between plane wakes, and its downstream location can be therefore modeled, for regular grids, as $x_\star \sim M^2/(C_d t)$, with $t$ being the thickness of the bars and $C_d$ their respective drag coefficient. The value of $x_\star$, that properly predicts the position of the fluctuations maximum for regular and fractal grids at large Reynolds number depends only on the geometry of the grid, and is independent of $U_\infty$.  

Conversely, our results with the OG hint that $x_\star$ is an increasing function of $U_\infty$ (figure \ref{fig5}d). This is an important difference with previous results in static grids, as it suggest an alternative turbulence generation mechanism not explained by the interaction between wakes, and possibly coming from interactions between laminar boundary layers or shear layers. To disentangle the underlying physics of this phenomenon, a broader range of $U_\infty$ should be explored (not currently possible with the present wind tunnel and grid) in conjunction with a particle image velocimetry (PIV) study near the grid.

\begin{figure}[t!]
\centering
\includegraphics[width=0.5\textwidth]{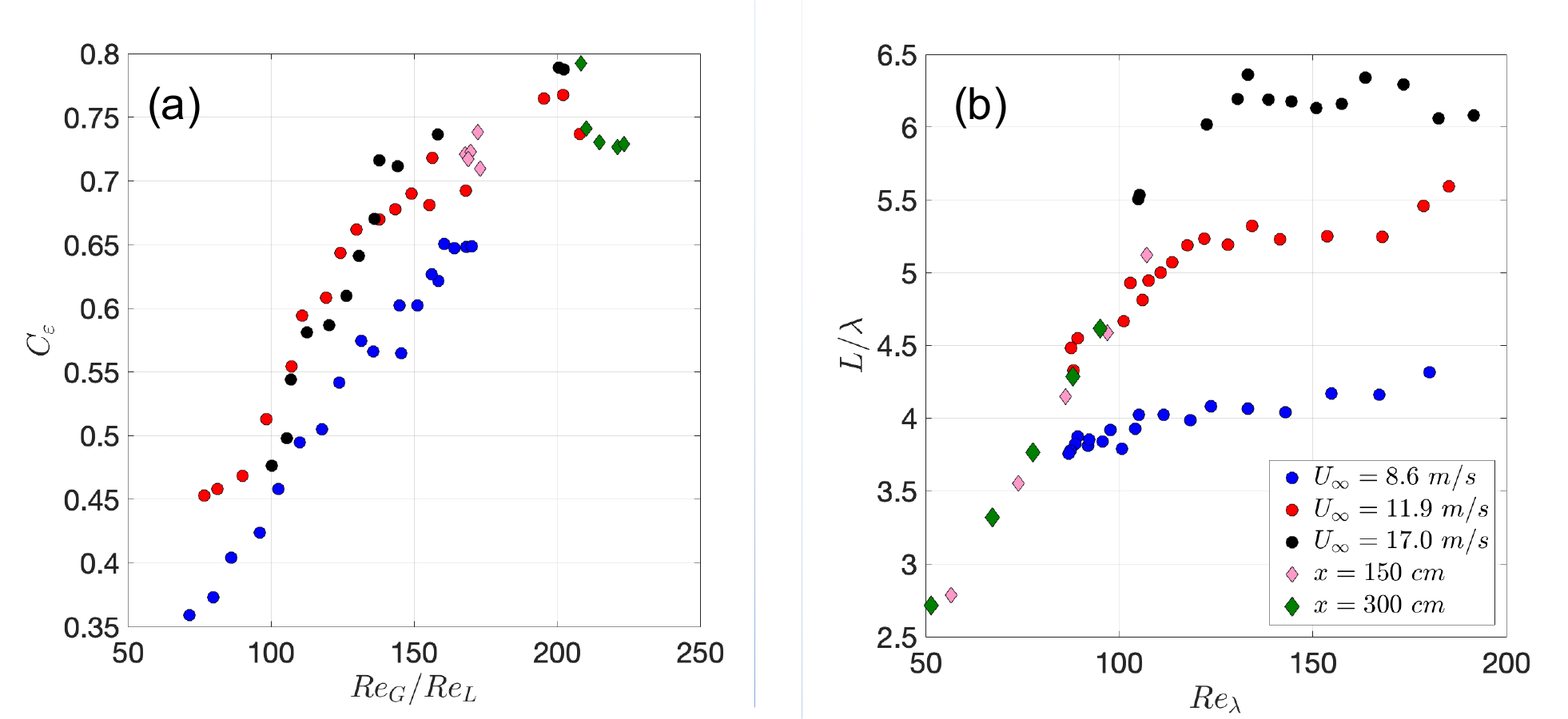}
\caption{(a) Plot of $C_\varepsilon$ vs  $Re_G/Re_L$. (b) $L/\lambda$ vs $Re_\lambda$. Results correspond to all datasets obtained for the open mode.}
\label{fig7}
\end{figure}

As the active mode only presents decaying turbulence, we will focus on the same regime for the open mode. We therefore report data only after the fluctuations peak. Figure \ref{fig7}a illustrates that $C_\varepsilon$ is a function that varies with $Re_G/Re_L$ (and ultimately with $Re_\lambda$). Furthermore, at low values of $Re_G/Re_L$ it collapses on a straight line, consistently with the high Reynolds non-equilibrium scalings. We also observe that $L/\lambda$ approaches a constant (figure \ref{fig7}b, the constant having a trend consistent with $\sqrt{Re_G}$) at large $Re_\lambda$ (i.e. low $Re_G/Re_L$). On the other hand at large $Re_G/Re_L$, we note that $C_\varepsilon$ becomes constant with $Re_\lambda$, and that $L/\lambda \sim C_\varepsilon Re_\lambda$, consistent with standard dissipation scalings. These figures, similar to those reported by Valente \& Vassilicos~\cite{valente2012universal} provide evidence  of the presence of both scalings (in separate regions of the flow), i.e., high Reynolds non-equilibrium scalings close downstream the kinetic energy peak, and standard ones after a transition region downstream. The outstanding fact of this transition (previously observed for fractal/regular static grids~\cite{valente2012universal} and in direct numerical simulations of periodic turbulence~\cite{goto2016unsteady}) is that occurs at relatively large values of $x$ (while in standard static grids happens at a few cm from it), allowing to capture both regimes in the same experiment. Moreover, the unfixed position of $x_\star$ implies that the open mode can be used to tailor the crossover downstream position between the two scalings. However, the presence of relatively large anisotropy values on the transition downstream position (visible in figure \ref{figiso}), suggests that the latter result may not be conclusive. Despite this, and given that the anisotropy barely changes in magnitude for different streamwise positions, the values of $C_\varepsilon$ estimated with the full kinetic energy instead of $u^\prime$ follow a similar trend.  Further studies on a larger wind tunnel (or with a grid with smaller $M$) may help to shed light on this phenomenon.

Up to now, our data suggests that the active grid, depending on the operating protocol selected, will produce different energy cascades, with important consequences for several applications. For instance, apart from the number of degrees of freedom and scale separation, the nature of the cascade will be related/affected by the persistence of coherent large scale structures, the turbulent kinetic energy budget and have consequences on terms of turbulence modelling~\cite{goto2016unsteady, vassilicos2015dissipation}. This reinforces the idea to develop an alternative method that allows to estimate and validate the values of $C_\varepsilon$ here obtained.

\subsection{Estimation of $\varepsilon$ via the zero-crossings of the streamwise velocity fluctuations}\label{sec:zeros}

Lieppman \cite{liepmann1953counting} was the first to adapt the Rice theorem so it can be applied to the zero-crossings of velocity fluctuations to estimate the Taylor micro-scale $\lambda$ in turbulent flows. In particular, he proved that $\lambda$ is proportional to the average distance $\bar{l}$ between zero-crossings points: $\bar{l}=B \lambda$, with $B=C \pi$ a constant that quantifies the non-Gaussianity of $\partial u / \partial x$ ($C=1$ for a Gaussian distribution, and $C>1$ for an intermittent turbulent flow). 

The theory that allows to deduce $C_\varepsilon$ from the number density of zero-crossings $n_s$ has been later developed~\cite{mazellier2008turbulence}, and relies on the fact that if the signal is low-pass filtered with a cut-off frequency $2\pi/\eta_c$ (being $\eta_c$ the scale of the filter not to be confused with the Kolmogorov lenghtscale $\eta$) , and that $n_s$ is a power law function of $L/\eta_c$ (as proposed by Sereenivasan and collaborators~\cite{sreenivasan1983zero} and by Davila \& Vassilicos~\cite{davila2003richardson}). Hence,

\begin{equation}
n_s=\frac{C_s^\prime}{L}(L/\eta_c)^{2/3},
\label{eq:ns}
\end{equation}

\noindent with $C_s'$ a dimensionless constant that characterizes the large scales. The  `$2/3$' exponent  is a consequence of the $-5/3$ power-law decay of the power spectral density of $u$~\cite{davila2003richardson}. 

\begin{figure}[t!]
\centering
\includegraphics[width=0.5\textwidth]{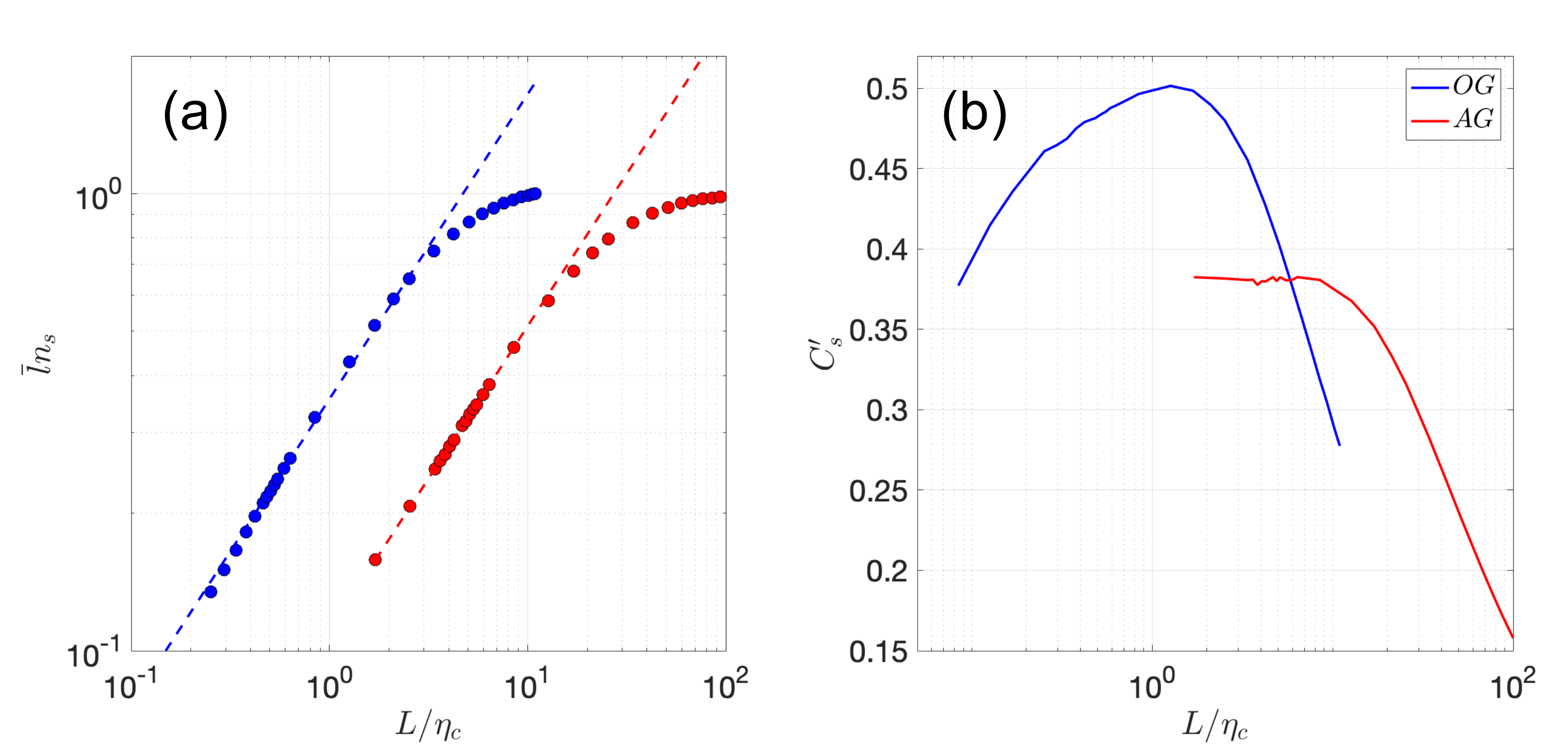}
\caption{$\bar{l} n_s$ vs $L/\eta_c$ (a). The dashed lines are a $2/3$ power law. $Cs'$ vs $L/\eta_c$ (b). Blue lines correspond to the open mode and the red ones to the active one.\label{fig8}}
\end{figure}

Using the previous equation, and the definition for the dissipation, and the Taylor lengthscale; $\varepsilon=C_\varepsilon \frac{u^{\prime3}}{L}$ and $\lambda=\sqrt{\frac{15 \nu u^{\prime2}}{\varepsilon}}$, the value of $C_\varepsilon$ can be computed as,

\begin{equation}\label{eqceps}
C_\varepsilon=(15B^2)^{3/2}\left(\frac{C_s\prime}{A^{2/3}}\right)^{3},
\end{equation}

\noindent with $A=\eta_*/ \eta$. $\eta_*$ is the value of $\eta_c$ that correspond to the intersection of the $2/3$ power law and the value $\bar{l}n_s=1$. At large values of $Re_\lambda$, it was found~\cite{mazellier2008turbulence} that the latter expression goes as $C_\varepsilon \sim C_s^{\prime3}$, which in turn implies that dissipation is controlled by the large scales of the flow. We remark that in any moment of the deduction it was used that $C_\varepsilon$ has to be constant (or dependent on $Re_G/Re_L$) and therefore equation \ref{eqceps} remains valid for any energy cascade (within only approximate HIT conditions, as Mazellier \& Vassilicos used the model even for measurements at the centreline of a round jet), and in particular for our results for both the open and active modes. 

To check the previous prediction, we first checked whether the number of zero crossings density $n_s$, followed a $2/3$ power law when $u$ is low-pass filtered with different cutoffs wavenumbers $2\pi/\eta_c$. Figure \ref{fig8}a shows that indeed the law was recovered by our dataset. Furthermore, figure \ref{fig8}b shows that $C_s^\prime$ is properly defined within the region that follows the power law. The value of $C_s^\prime$ presents a maximum instead of a plateau for the open mode, a consequence of the low values of $Re_\lambda$ (also observed in previous regular grid measurements at similar values of $Re_\lambda$~\cite{mazellier2008turbulence}). Therefore, we took the maximum values of $C_s^\prime$ for the open mode. 

The results reported here correspond to an antialiasing finite impulse response lowpass filter. We found that the parameter $\eta_*$ depends on the properties and type of filter used, affecting both the values of $A$ and $C_s^\prime$. Nevertheless, for all cases, the values of $C_\varepsilon$ deduced are very robust (as the dependency with $\eta_*$ is cancelled on equation \ref{eqceps}).

\begin{figure}[t!]
\centering
\includegraphics[width=0.5\textwidth]{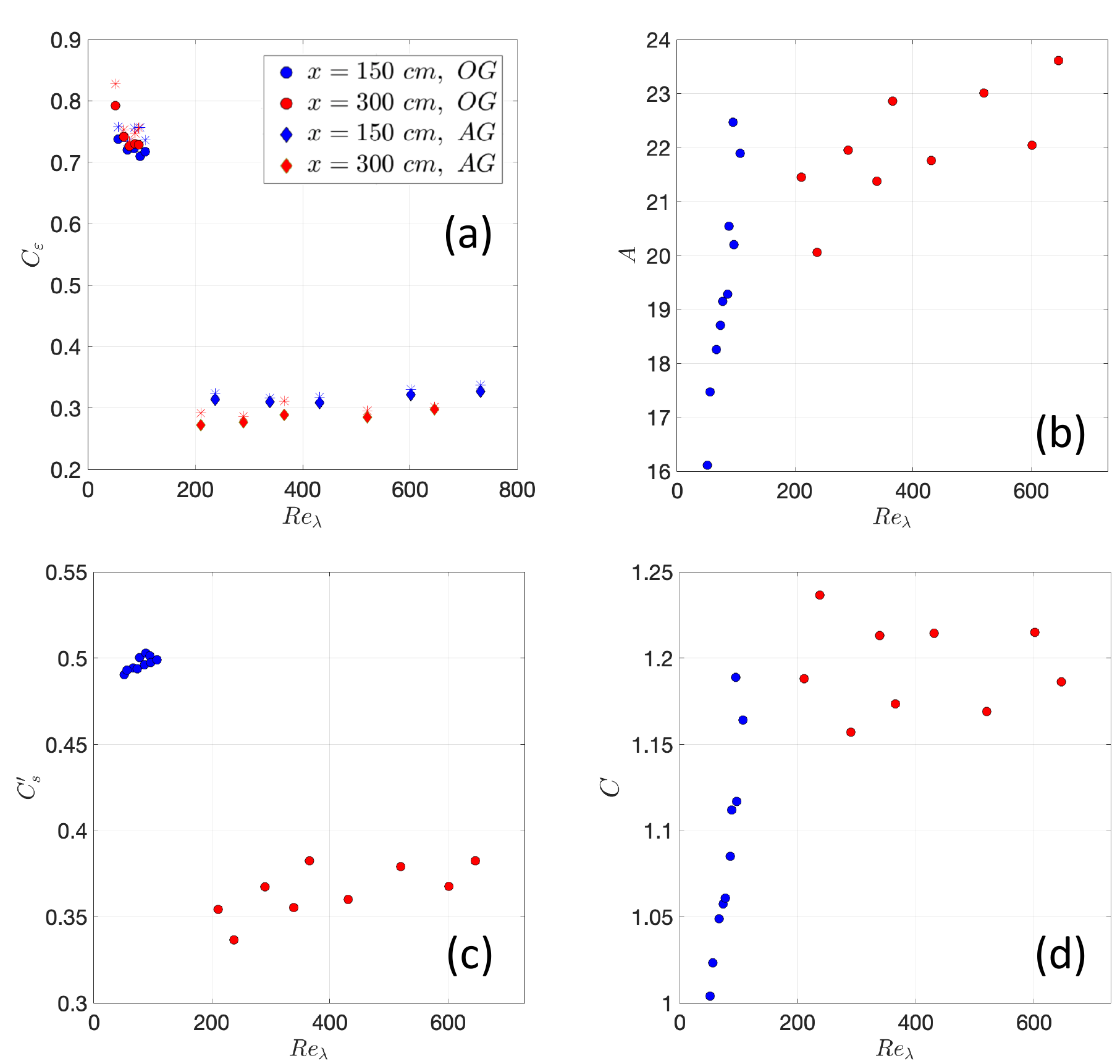}
\caption{(a) Comparison of the values $C_\varepsilon^{model}$ obtained using equation \ref{eqceps} (star symbols) and the experimental ones $C_\varepsilon^{exp}=\varepsilon L / u^{\prime3}$ (filled symbols). Value of $A$ (b), $C_s^\prime$ (c) and the intermittency constant $C$ (d) vs $Re_\lambda$ for all datasets.}
\label{fig9}
\end{figure}

Once these requirements were verified, the validity of equation \ref{eqceps} can be assessed. The value of $C$ could be, in principle deduced from the velocity derivatives. However, as the active mode has large values of $Re_\lambda$, the convergence of the pdf of $\partial u / \partial x$ was not completely achieved. Instead, we took this value as $C=\frac{\bar{l}}{\lambda \pi}$.

Figure \ref{fig9}a shows streamwise profiles of $C_\varepsilon$ for our experimental data, and computed from from $C_\varepsilon^{exp}=\varepsilon L / u^{\prime3}$ and from equation \ref{eqceps} ($C_\varepsilon^{model}$) for every inflow conditions. The model predictions have good agreement with all values for the whole range of $Re_\lambda$. This is a remarkable result, considering it is the first observation of the 
validity of this model on non-equilibrium turbulence, and more importantly, it confirms the consistency of the values $C_\varepsilon^{exp}$ here obtained.

Figures \ref{fig9}b-d show the values of $A$, $C_s^\prime$ and $C$, respectively. It can be clearly seen that these values strongly depend on the operating mode (OG/AG). The intermittency (indirectly quantified by $C$) increases with $Re_\lambda$ and seems to reach a constant value of $C\sim1.2$ for the active grid mode. As $A$ and $B$ tend to a constant at large $Re_\lambda$, we confirm the prediction that $C_\varepsilon \sim C_s^{\prime3}$ for large $Re_\lambda$. 

\section{Conclusions}

We present a systematic study of the streamwise evolution of active-grid-generated turbulence for two paradigmatic operating modes; a high $Re_\lambda$ active and random, and a low $Re_\lambda$ static one. We found the first evidence that the energy cascade may be strongly influenced by the operating protocol. This outcome has important consequences for active grids related research, e.g., the interaction of wind turbines with background turbulence, the formation and development of clusters of inertial particles, among many others that will rely on the nature of a given energy cascade.
 
We corroborated previous experimental measurements, which showed $C_\varepsilon$ is constant for active-grid turbulence in the triple random mode (and therefore consistent with both the Richardson-Kolmogorov and a balanced non-equilibrium energy cascades). On the other hand, we observed that on a region close downstream the turbulent kinetic energy peak, the open mode followed recently proposed high Reynolds non-equilibrium scalings. Furthermore, this mode seems to be generating a particular type of turbulence (not observed before), which is possibly controlled by the boundary and/or shear layers at the blades of the grid, as the -so defined- wake interaction length is an increasing function of the inlet velocity, and not only dependent on the grid geometry. 

Furthermore, we find that the mesh size of the grid is no longer a good estimation of $L$ for any of the tested operating modes. The active mode produces integral length scales in the order of the measurement section (as proposed already on~\cite{makita1991active}). On the other hand, the open mode has an integral scale of around 1/5 the mesh size. 

Further studies for intermediate protocols, between the static and random ones, could shed light on underlying mechanics behind the transition from standard to non-equilibrium cascades. Our results seem to agree with the suggestion that non-equilibrium turbulence is related to the presence of large-scale coherent structures~\cite{goto2016unsteady}, as the active mode may destroy them very fast while it could be expected that they are more persistent for the open one. Interestingly, in this work the non-equilibrium energy cascade occurs for the flow at globally lower $Re_\lambda$, while for a fixed grid geometry, these type of cascade occur close downstream to the grid (and therefore where $Re_\lambda$ is larger). This is consistent with the findings from~\cite{goto2016unsteady}: non-equilibrium turbulence do not seem to be a phenomenon that depend uniquely on the Reynolds number of the flow.

Finally, we verified that the model developed by Mazellier \& Vassilicos~\cite{mazellier2008turbulence} extends to non-equilibrium scalings. This theoretical model, that predicts the value of $C_\varepsilon$ from the zero-crossings of velocity fluctuations, is a powerful tool to assess the validity of the results obtained when different operating protocols are employed.

\begin{acknowledgments}
We thank Sebastien Torre for his help with installing and operating the traverse system. This work has been partially supported by the LabEx Tec21 (Investissements d’Avenir - Grant Agreement \# ANR-11-LABX-0030). 
\end{acknowledgments}

\bibliographystyle{plain}

\end{document}